\documentstyle[twocolumn]{mn}
\newcommand{\mincir}{\raise
-2.truept\hbox{\rlap{\hbox{$\sim$}}\raise5.truept
\hbox{$<$}\ }}
\newcommand{\magcir}{\raise
-2.truept\hbox{\rlap{\hbox{$\sim$}}\raise5.truept
\hbox{$>$}\ }}
\newcommand{\minmag}{\raise-2.truept\hbox{\rlap{\hbox{$<$}}\raise
6.truept\hbox
{$>$}\ }}
\newcommand{\be}{\begin{equation}}
\newcommand{\ee}{\end{equation}}
\newcommand{\ba}{\begin{eqnarray}}
\newcommand{\ea}{\end{eqnarray}}
\newcommand{\brr}{\begin{array}}
\newcommand{\nn}{\nonumber \\} 
\newcommand{\err}{\end{array}}
\newcommand{\bc}{\begin{center}}
\newcommand{\ec}{\end{center}}

\input{psfig.sty} 
\title[Clustering 
properties of galaxy clusters in optical and X-ray bands]
{Constraining cosmological parameters with the clustering 
properties of galaxy clusters in optical and X-ray bands}
\author[Moscardini et al.]
{L. Moscardini$^1$, S. Matarrese$^{2,3}$ and H.J. Mo$^3$ \\ 
$^1$Dipartimento di Astronomia, Universit\`a di
Padova, vicolo dell'Osservatorio 2, I--35122 Padova, Italy\\
$^2$Dipartimento di Fisica G. Galilei, Universit\`{a} di Padova, via
Marzolo 8, I--35131 Padova, Italy\\ 
$^3$Max-Planck-Institut f\"ur Astrophysik, Karl-Schwarzschild-Strasse
1, D-85748 Garching, Germany }

\begin{document}

\maketitle

\begin{abstract}
We use a theoretical model to predict the clustering properties of
galaxy clusters. Our technique accounts for past light-cone effects on
the observed clustering and follows the non-linear evolution in
redshift of the underlying dark matter correlation function and
cluster bias factor.  A linear treatment of redshift-space distortions
is also included.  We perform a maximum-likelihood analysis by
comparing the theoretical predictions to a set of observational data,
both in the optical (two different subsamples of the APM catalogue and
the EDCC catalogue) and X-ray band (RASS1 Bright Sample, BCS, XBACs,
REFLEX). In the framework of cold dark matter models, we compute the
constraints on cosmological parameters, such as the matter density
$\Omega_{\rm 0m}$, the cosmological constant $\Omega_{\rm 0\Lambda}$,
the power-spectrum shape parameter $\Gamma$ and normalisation
$\sigma_8$.  Our results show that X-ray data are more powerful than
optical ones, allowing smaller regions in the parameter space.  If we
fix $\Gamma$ and $\sigma_8$ to the values suggested by different
observational datasets, we obtain strong constraints on the matter
density parameter: $\Omega_{\rm 0m}\le 0.5$ and $0.2\le
\Omega_{\rm 0m}\le 0.35$, for the optical and X-ray data, respectively.
Allowing the shape parameter to vary, we find that the clustering
properties of clusters are almost independent of the matter density
parameter and of the presence of a cosmological constant, while they
appear to be strongly dependent on the shape parameter.
Using the X-ray data only, we obtain $\Gamma\sim 0.1$ and $0.4\mincir
\sigma_8 \mincir 1.1$ for the Einstein-de Sitter model, while $0.14 \mincir
\Gamma \mincir 0.22$ and $0.6 \mincir \sigma_8 \mincir 1.3$ for open 
and flat models with $\Omega_{\rm 0m}=0.3$.  Finally, we use our model
to make predictions on the correlation length of galaxy clusters
expected in future surveys. In particular, we show the results for an
optical catalogue with characteristics similar to the EIS project and
for a very deep X-ray catalogue with the characteristics of the
XMM/LSS survey. We find that clusters at high redshifts are expected
to have larger a correlation length than local ones.
\end{abstract}

\begin{keywords}
cosmology: theory -- galaxies: clusters -- large--scale structure of
Universe -- X-rays: galaxies -- dark matter
 
\end{keywords}

\section{Introduction}

Clusters of galaxies are the largest collapsed objects with masses
dominated by dark matter. Their properties are largely determined by
the gravitational collapse in the cosmological density field, and so
their distribution in the Universe depends both on cosmology and on
the initial density perturbations. Therefore, the study of the
clustering of clusters may provide important diagnostics for models of
structure formation. One important advantage of using clusters to
study the large-scale structure is that their formation in the cosmic
density field is relatively easy to understand. Indeed, modern
cosmological simulations show that clusters of galaxies may be
identified as the most massive dark haloes produced by gravitational
collapse, and so their clustering properties are relatively easy to
interpret theoretically. In fact, based on an extension of the
Press-Schechter formalism, one can obtain an analytical model in which
the dependence of the two-point correlation function of dark haloes on
cosmology and initial power spectrum can be explicitly seen (Mo \&
White 1996; Catelan et al. 1998; Sheth, Mo \& Tormen 2001).

It has been known for some time that the strong observed correlation
of clusters on large scales is very difficult to reconcile with the
standard cold dark matter (CDM) model (White et al. 1987; Dalton et
al. 1992; Jing et al. 1993; Mo, Peacock \& Xia 1993; Dalton et
al. 1994; Borgani, Coles \& Moscardini 1994; Croft \& Efstathiou 1994;
Borgani et al. 1995). More recently, analyses on the constraints by
the cluster-cluster correlation function on current theoretical models
of structure formation have been carried out by a number of authors
(Eke et al. 1996; Mo, Jing \& White 1996; Moscardini et al. 2000a,b;
Robinson 2000; Colberg et al. 2000).

In this paper, we use current available observational data on the
cluster two-point correlation function to constrain theoretical models
of structure formation. We use analytical tools well tested by
numerical simulations to make predictions for a large grid of
models. Our analysis follows closely that of Mo et al. (1996) and
Robinson (2000).  However, there are several important
differences. First, we consider new datasets; in particular, we use
data from recent X-ray surveys (see also Moscardini et al. 2000b).
Second, we use improved theoretical models (Matarrese et al. 1997).
In particular we pay much attention to the redshift evolution in the
clustering of clusters and to light-cone and selection effects.
Third, we present theoretical predictions for some future surveys.
  
The plan of the paper is as follows. In Section 2 we list the
observational cluster samples used in the following analysis. Section
3 is devoted to the presentation of our theoretical model to estimate
the correlation of galaxy clusters both in the optical and in the
X-ray bands.  In Section 4 we present the results of a
maximum-likelihood analysis performed on existing data and we discuss
the constraints on the cosmological parameters. In Section 5 we show
the predictions of the correlation length for possible future
surveys. Conclusions are drawn in Section 6.

\section{Cluster samples}

In this section we present the observational data on the galaxy
cluster correlation length that we will use in the following analysis
to constrain the cosmological parameters. A summary of these data is
presented in Table \ref{t:data}.

\begin{table*}
\centering
\caption[]{The clustering data (in the optical and X-ray bands) used
in the likelihood analysis. Column 1: catalogue name. Column 2:
characteristics of the catalogue: for the optical catalogues, the
richness ${\cal R}$ and the mean intracluster distance $d_c$; for the
X-ray catalogues, the limiting luminosity $L_X$ or the limiting flux
$S_{\rm lim}$, unless the whole catalogue is analyzed.  Column 3:
reference for the clustering analysis.  Column 4: number of clusters
in the catalogue $n_c$. Columns 5, 6 and 7: correlation length $r_0$
(in $h^{-1}$ Mpc), the slope of the correlation function $\gamma$ and
the corresponding confidence levels of the quoted errorbars.}
\tabcolsep 4pt
\begin{tabular}{lllllll} \\ \\ \hline \hline
Catalogue & Characteristics & Reference & $n_c$ & $r_0$ ($h^{-1}$ Mpc)
& $\gamma$ & Errorbars \\ 
\hline 
Optical band: & & & & & & \\ 
APM Sample B& ${\cal R}\ge 50$, $d_c=30$ $h^{-1}$ Mpc&Croft et al. (1997)& 
$364$ & $14.2^{+0.4}_{-0.6}$ & $2.13^{+0.09}_{-0.06}$ & $1\sigma$\\ 
APM Sample C& ${\cal R}\ge 80$, $d_c=57$ $h^{-1}$ Mpc&{\it idem}& 
$110$ & $18.4^{+2.2}_{-2.4}$ & $ 1.7^{+0.3}_{-0.3}$ & $1\sigma$\\ 
EDCC & $d_c=46$ $h^{-1}$ Mpc&Nichol et al. (1992) & 
$79$ & $16.4\pm 4.0$ & $2.1\pm 0.3 $ & $1\sigma$\\
\hline
X-ray band: & & & & & & \\
RASS1 Bright Sample &  whole catalogue &Moscardini et al. (2000a) & 130
&
$21.5^{+3.4}_{-4.4}$ &  $2.11^{+0.53}_{-0.56}$ &
$2\sigma$ \\
BCS& $L_X\ge
0.24\times 10^{44} h^{-2}$ erg s$^{-1}$ &Lee \& Park (1999)&33&
$33.0^{+6.2}_{-5.9}$ & $1.82^{+0.49}_{-0.50}$ &
$1\sigma$\\
XBACs &   whole catalogue& Borgani et al. (1999) &203  &  $26.0 \pm 4.5$
&  $1.98^{+0.35}_{-0.53}$ &
$2\sigma$ \\ 
REFLEX & $S_{\rm lim}=3\times 10^{-12}$ erg cm$^{-2}$ s$^{-1}$ & Collins
et al. (2000) &449& $18.8\pm 0.9$ &  $1.83^{+0.15}_{-0.08}$ &
$1\sigma$ \\
\hline
\end{tabular}
\label{t:data}
\end{table*}

\subsection{Optical data}

As far as the optical band is concerned, we will use for our study the
observational results from the analysis by Croft et al. (1997) on the
APM cluster redshift survey and by Nichol et al. (1992) on the
Edinburgh-Durham cluster catalogue (EDCC).

\begin{itemize}
\item
To construct the APM cluster survey, an automated procedure has been
used to select from the angular APM galaxy survey the clusters whose
redshifts were later measured.  An early sample, containing 364
clusters with richness ${\cal R}\ge 50$, called Sample B (Dalton et
al. 1994), has been created by fixing the limiting magnitude of the
galaxy catalogue to $b_j=20.5$. A new, richer sample was then obtained
by Croft et al. (1997) by extending the limit to $b_j=21.0$ and
considering only clusters with richness ${\cal R}\ge 80$. This
catalogue (called Sample C) contains 165 objects with redshift $cz\le
55000 $ km s$^{-1}$.  Even if the clustering study made by Croft et
al. (1997), using a maximum-likelihood technique, considered six
different subsamples with decreasing cluster density, in our
statistical analysis we will prefer to use the results only for the
two whole samples B and C. This is done to reduce the possible
problems coming from the assumption of independence of the different
samples. This assumption is required by the way we write the
likelihood (see equation \ref{eq:chi2}). Our final results will not be
affected by this choice, as shown by the fact that the confidence
levels on the cosmological parameters obtained using sample B only are
almost indistinguishable from those coming from the complete analysis
of the optical datasets.  The corresponding mean intracluster
separations for samples B and C are 30 and 57 $h^{-1}$ Mpc,
respectively ($h$ is the value of the local Hubble constant $H_0$ in
units of 100 km s$^{-1}$ Mpc$^{-1}$); the values of the correlation
lengths $r_0$, of the slope $\gamma$ of the correlation function, and
their $1\sigma$ errorbars, obtained from a maximum likelihood
analysis, are reported in Table \ref{t:data}.

\item
The EDCC is a machine-based, objectively selected sample of galaxy
clusters consisting of 737 objects of all richness, over 0.5 sr of sky
centred on the South Galactic Pole (Lumsden et al. 1992). A subsample
of 79 clusters, with at least 22 galaxies inside a radius
corresponding to 1 $h^{-1}$ Mpc, with magnitudes between the limits
$m_3$ and $m_3+2$ (see Abell 1958 for the definitions) has been
spectroscopically confirmed. The resulting catalogue, with a
corresponding mean intracluster separation $d_c=46 h^{-1}$ Mpc, has
been used by Nichol et al. (1992) to estimate the clustering
properties.  The values of $r_0$ and $\gamma$ obtained from a
least-squares fit are reported in Table \ref{t:data}.
\end{itemize}

\subsection{X-ray data}

In the following analysis we will consider four different catalogues
of galaxy clusters selected in the X-ray band: the $ROSAT$ All-Sky
Survey 1 (RASS1) Bright Sample, the $ROSAT$ Brightest Cluster Sample
(BCS), the X-ray brightest Abell-type cluster sample (XBACs), the
$ROSAT-ESO$ Flux-limited X-ray sample (REFLEX).  Here we list some of
their characteristics; more details can be found in the original
papers.

\begin{itemize}
\item
The RASS1 Bright Sample (De Grandi et al. 1999a) contains 130 clusters
of galaxies selected from the $ROSAT$ All-Sky Survey (RASS) data.  The
catalogue has an effective flux limit in the (0.5 -- 2.0 keV) band
between 3.05 and $4\times 10^{-12}$ erg cm$^{-2}$ s$^{-1}$ over the
selected area which covers a region of approximately 2.5 sr within the
Southern Galactic Cap, i.e. $\delta<2.5^o$ and $b<-20^o$.  For our
theoretical predictions we will use the exact sky map of the sample
which is presented in Figure 2 of De Grandi et al. (1999a). The
redshift distribution has a tail up to $z\simeq 0.3$ but the majority
of the clusters have $z<0.1$. Moscardini et al. (2000a) found that the
two-point correlation function of the whole sample is well fitted by a
power-law with $r_0=21.5^{+3.4}_{-4.4} h^{-1}$ Mpc and $\gamma =
2.11^{+0.53}_{-0.56}$ (95.4 per cent confidence level with one fitting
parameter).

\item
The BCS catalogue (Ebeling et al. 1998) is an X-ray selected,
flux-limited sample of 201 galaxy clusters with $z\le 0.3$ drawn from
the RASS data in the northern hemisphere ($\delta \ge 0^o$) and at
high Galactic latitude ($|b_{II}|\ge 20^o$).  The limiting flux is
$S_{\rm lim}=4.45 \times 10^{-12}$ erg cm$^{-2}$ s$^{-1}$ in the
(0.1--2.4) keV band. Since the sky-coverage $\Omega_{\rm sky}(S)$ of
BCS is not available, we will use $\Omega_{\rm sky}(S) = {\rm const}
\simeq 4.13 $ steradians for fluxes larger than $S_{\rm lim}$.  Lee \&
Park (1999) analyzed the clustering properties of this catalogue using
four different volume-limited subsamples. For simplicity, in the
following analysis we will consider only the catalogue with $L_X\ge
0.24\times 10^{44} h^{-2}$ erg s$^{-1}$ and a limiting redshift of
$z=0.07$.  The correlation function in this case is fitted by a
power-law with $r_0=33.0^{+6.2}_{-5.9} h^{-1}$ Mpc and $\gamma =
1.82^{+0.49}_{-0.50}$ (68.3 per cent confidence level with one fitting
parameter). We checked that the inclusion of the deeper subsamples
produced very small changes in our likelihood analysis.

\item
The XBACs catalogue (Ebeling et al. 1996) is an all-sky X-ray sample
of 242 Abell galaxy clusters extracted from the RASS data. Being
optically selected, it is not a complete flux-limited catalogue. The
sample covers high Galactic latitudes ($|b_{II}|\ge 20^o$). The
adopted limiting flux is $S_{\rm lim}=5 \times 10^{-12}$ erg cm$^{-2}$
s$^{-1}$ in the (0.1--2.4) keV band.  Also in this case, since the
actual sky coverage is not available, we will adopt $\Omega_{\rm
sky}(S)= {\rm const} \simeq 8.27 $ steradians for fluxes larger than
$S_{\rm lim}$.  The aforementioned selection effects produce a
luminosity function for XBACs which is much lower in the faint part
than that obtained from other catalogues (e.g. Ebeling et al. 1997;
Rosati et al. 1998; De Grandi et al. 1999b).  We take into account
this incompleteness in our model following the same method described
in Moscardini et al. (2000b).  The clustering properties of this
catalogue have been studied by different authors. Abadi, Lambas \&
Muriel (1998) found for the whole catalogue a correlation length
$r_0=21.1^{+1.6}_{-2.3} h^{-1}$ Mpc (1$\sigma$ errorbar) and a slope
$\gamma=-1.92$. Borgani, Plionis \& Kolokotronis (1999) analyzed the
same sample finding a somewhat larger correlation amplitude:
$r_0=26.0^{+4.1}_{-4.7} h^{-1}$ Mpc and $\gamma =
1.98^{+0.35}_{-0.53}$ (95.4 per cent confidence level with one fitting
parameter).  The difference between these two estimates is probably
due to a different assumption on the errors. Since the
maximum-likelihood approach is more robust than the quasi-Poisson
assumption made by Abadi et al. (1998), we prefer to use in the
following analysis the results obtained by Borgani et al. (1999).

\item
The REFLEX survey (B\"ohringer et al. 1998) is a large sample of
optically confirmed X-ray clusters selected from RASS.  The sample
includes 452 X-ray selected clusters in the southern hemisphere, at
high Galactic latitude ($|b_{II}|\ge 20^o$). For our computations, we
use the actual sky coverage given in Figure 1 of Collins et
al. (2000). Using a catalogue with a limiting flux of $3 \times
10^{-12}$ erg cm$^{-2}$ s$^{-1}$ (defined in the 0.1 -- 2.4 keV energy
band), where the sky coverage falls to 97.4 per cent of the whole
surveyed region (4.24 steradians), Collins et al. (2000) found that
the two-point correlation function is fitted by a power-law with
$r_0=18.8\pm 0.9 h^{-1}$ Mpc and $\gamma = 1.83^{+0.15}_{-0.08}$ (68.3
per cent confidence level from a maximum-likelihood analysis).  As for
the APM subsamples at higher richness, in our analysis we prefer not
to use the clustering results for the REFLEX subsamples at higher
luminosities, also discussed in Collins et al. (2000).

\end{itemize}

\section{Theoretical models for the correlation function}

Our theoretical predictions for the spatial two-point correlation
function of galaxy clusters in different cosmological models have been
here obtained by using an updated version of the method presented in
Moscardini et al. (2000a,b), where the application was limited to
X-ray selected clusters.  Here we will give only a short description
of the method and we refer to those papers for a more detailed
discussion.

\subsection{Clustering in the past-light cone}

Matarrese et al. (1997; see also Moscardini et al. 1998 and Yamamoto
\& Suto 1999 and references therein) developed an algorithm to
describe the clustering in our past light-cone taking into account
both the non-linear dynamics of the dark matter distribution and the
redshift evolution of the bias factor.  The final expression for the
observed spatial correlation function $\xi_{\rm obs}$ in a given
redshift interval ${\cal Z}$ is
\be 
\xi_{\rm obs}(r) = { \int_{\cal Z}
d z_1 d z_2 {\overline{\cal N}}(z_1) {\overline{\cal N}}(z_2)
~\xi_{\rm obj}(r;z_1,z_2) \over \bigl[ \int_{\cal Z} d z_1
{\overline{\cal
N}}(z_1) \bigr]^2 } \;,
\label{eq:xifund}
\ee 
where ${\overline{\cal N}}(z)\equiv {\cal N}(z)/r(z)$ and ${\cal
N}(z)$ is the actual redshift distribution of the catalogue. In the
previous formula $\xi_{\rm obj}(r,z_1,z_2)$ represents the correlation
function of pairs of objects at redshifts $z_1$ and $z_2$ with
comoving separation $r$. An accurate approximation for it is given by
\be 
\xi_{\rm obj}(r,z_1,z_2) \approx b_{\rm
eff}(z_1) b_{\rm eff}(z_2) \xi_{\rm m}(r,z_{\rm ave}) \;, 
\ee 
where $\xi_{\rm m}$ is the dark matter covariance function and $z_{\rm
ave}$ is a suitably defined intermediate redshift.

Another important ingredient entering the previous equation is the
effective bias $b_{\rm eff}$ which can be expressed as a weighted
average of the `monochromatic' bias factor $b(M,z)$ of objects of some
given intrinsic property $M$ (like mass, luminosity, etc):
\be 
b_{\rm eff}(z) \equiv {\cal N}(z)^{-1} \int_{\cal M} d\ln M' ~b(M',z) 
~{\cal N}(z,M')\, ,
\label{eq:b_eff}
\ee
where ${\cal N}(z,M)$ is the number of objects actually present in the
catalogue with redshift in the range $z,~z+dz$ and $M$ in the range
$M,~M+dM$, whose integral over $\ln M$ is ${\cal N}(z)$.

As galaxy clusters are expected to form by the hierarchical merging of
smaller mass units, one can fully characterize their properties by the
mass $M$ of their hosting dark matter haloes at each redshift $z$. One
can then estimate their comoving mass function $\bar n(z,M)$ by the
Press-Schechter (1974) formula and adopt the Mo \& White (1996)
relation for the monochromatic bias.  Actually, we use the relations
recently introduced by Sheth \& Tormen (1999) and Sheth, Mo \& Tormen
(2001), which have been shown to produce a more accurate fit of the
distribution of the halo populations in numerical simulations (Jenkins
et al. 2001).

To complete our model we need a technique to compute the redshift
evolution of the dark matter covariance function $\xi_{\rm m}$. For
this, we use the fitting formula by Peacock \& Dodds (1996), which
allows to evolve $\xi_{\rm m}$ into the fully non-linear regime.

\subsection{From the catalogue characteristics to the halo mass}

In order to predict the abundance and clustering of galaxy clusters in
the different catalogues described above (both in the optical and
X-ray bands) we need to relate the sample characteristics to a
corresponding halo mass at each redshift.

For the optical catalogues, the various samples of APM and EDCC
considered by Croft et al. (1997) and Nichol et al. (1992)
respectively, are characterized by a different cluster number density
(see Table \ref{t:data}). In this case, for each cluster subsample, we
fix the minimum mass (which in our model represents the lower limit of
the integral in equation \ref{eq:b_eff}) of the hosting dark matter
haloes in such a way that the comoving cumulative mass function
reproduces the number density of clusters in the range of redshift
sampled by the original catalogues.

For X-ray selected clusters, we use a different approach (see
Moscardini et al. 2000a,b and Suto et al. 2000).  In this case it is
quite easy to relate the limiting flux (and/or eventually the limiting
luminosity, as required for instance by the BCS subsample that we
consider) to the minimum halo mass.  In fact the flux $S$ in a given
band corresponds to an X-ray luminosity $L_X=4\pi d_L^2 S$ in the same
band, where $d_L$ is the luminosity distance.  The quantity $L_X$ can
be converted into the total luminosity $L_{\rm bol}$ by performing
band and bolometric corrections (we assume an overall ICM metallicity
of $0.3$ times solar).  Local observations suggest that the cluster
bolometric luminosity is related to the temperature by a simple
relation: $T = {\cal A} \ L_{\rm bol}^{\cal B}$, where the temperature
is expressed in keV and $L_{\rm bol}$ is in units of $10^{44} h^{-2}$
erg s$^{-1}$.  We assume ${\cal A}=4.2$ and ${\cal B}=1/3$, which are
a good representation of the data with $T\magcir 1$ keV
(e.g. Markevitch 1998 and references therein). Below this temperature
(i.e. for galaxy groups) the $L_{\rm bol}-T$ relation has much steeper
a slope: for this reason we set a minimum temperature at $T=1$ keV.
Recent analyses of cluster temperature data at higher redshifts
(Mushotzky \& Scharf 1997; Donahue et al. 1999) are consistent with no
evolution in the $L_{\rm bol}-T$ relation out to $z \approx
0.4$. Therefore, we can safely assume that the previous relation holds
true in the redshift range sampled by the considered catalogues.
Moscardini et al. (2000b), who allowed a moderate redshift evolution
of the $L_{\rm bol}-T$ to reproduce the observed $\log N$--$\log S$
relation, showed that the clustering properties are only slightly
sensitive to this assumption.  To convert the cluster temperature into
the mass of the hosting dark matter halo, we assume virial isothermal
gas distribution and spherical collapse: $T \propto M^{2/3} E^{2/3}(z)
\Delta_{\rm vir}(z)^{1/3}$, where $\Delta_{\rm vir}$ represents the
mean density of the virialized halo in units of the critical density
at that redshift and $E(z)$ is the ratio between the value of the
Hubble constant at redshift $z$ and today.  Once the relation between
observed flux and halo mass at each redshift is established we have to
account for the catalogue sky coverage $\Omega_{\rm sky}(S)$ (when
available) to predict the redshift distribution.

\subsection{Redshift distortions}

We include the effect of redshift-space distortions using linear
theory and the distant-observer approximation (Kaiser 1987).  Under
these assumptions the enhancement of the redshift-space averaged power
spectrum is approximately $1+2\beta/3+\beta^2/5$, where $\beta\simeq
\Omega_{\rm 0m}^{0.6}/b_{\rm eff}$.  We find that the total effect of
redshift distortions on the value of the correlation length is always
smaller than 10 per cent.

\section{Results}

\subsection{Cosmological models}

In what follows we use a maximum-likelihood analysis to constrain the
main parameters defining a cosmological model.  We will consider a set
of structure formation models all belonging to the general class of
cold dark matter (CDM) ones, for which the linear power-spectrum reads
$P_{\rm lin}(k,0) \propto k^n T^2(k)$, with $T(k)$ the CDM transfer
function (Bardeen et al. 1986)
\ba
T(q) & = & {\ln \left( {1+2.34q}\right) \over 2.34q}\ \times 
[1+3.89q+ \nn
& & (16.1q)^2+(5.46q)^3+(6.71q)^4]^{-1/4}\ ,
\ea
where $q=k/h\Gamma$. The shape parameter $\Gamma$ depends on the
Hubble parameter $h$, on the matter density $\Omega_{\rm 0m}$ and on
the baryon density $\Omega_{\rm 0b}$ (Sugiyama 1995):
\be 
\Gamma\ =\ \Omega_{\rm 0m} h\exp (-\Omega_{\rm 0b}-\sqrt{h/0.5}\
\Omega_{\rm 0b}/\Omega_{\rm 0m})\ .  
\ee

We fix the spectral index $n$ to unity and we allow $\Gamma$ to vary
in the range 0.05--0.5, while $\Omega_{\rm 0m}$ ranges from 0.1 to 1
in the framework of both open and flat models, with a cosmological
constant contributing to the total density with $\Omega_{\rm
0\Lambda}=1-\Omega_{\rm 0m}$. Finally, we use different normalisations
of the primordial power-spectrum, parameterized by $\sigma_8$ (the
r.m.s. fluctuation amplitude in a sphere of $8 h^{-1}$ Mpc) in the
range $0.2\le \sigma_8 \le 2$.

In summary, the cosmological models we consider are defined by four
parameters: $\Omega_{\rm 0m}$, $\Omega_{\rm 0\Lambda}$, $\Gamma$ and
$\sigma_8$.

\subsection{Maximum likelihood  analysis}

Confidence levels for the cosmological parameters are obtained through
a maximum likelihood analysis.  The likelihood is ${\cal L} \propto
\exp(-\chi^2/2)$, where
\be
\chi^2=\sum_{i=1}^{N_{\rm data}} {{[r_0(i)-r_0(i; \Omega_{\rm 0m},
\Omega_{\rm 0\Lambda},
\Gamma, \sigma_8)]^2}\over {\sigma^2_{r_0}(i)}}\ .
\label{eq:chi2}
\ee 
The sum runs over the observational dataset described in Section 2,
i.e. $N_{\rm data}=3$ and $N_{\rm data}=4$ for the optical and X-ray
bands, respectively.  The quantities $r_0(i)$ and $\sigma_{r_0}(i)$
represent the values of the correlation length and its $1\sigma$
errorbar for each catalogue, as reported in Table \ref{t:data};
$r_0(i; \Omega_{\rm 0m}, \Omega_{\rm 0\Lambda}, \Gamma, \sigma_8)$ is
the corresponding theoretical prediction obtained with a given choice
of cosmological parameters.  Note that the observed correlation
lengths reported in Table \ref{t:data} have been obtained using
different fitting methods (least-squares or maximum-likelihood
analyses) and over different spatial scales in ways which could be
inconsistent with each other. Moreover, the quoted values of $r_0$
come from fits of the data where also the values of the slope are
marginalized and it is known that there is significant covariance
between $r_0$ and $\gamma$. In spite of this, we prefer not to
estimate $r_0$ through the standard power-law fit procedure with two
parameters, because in the case of the maximum-likelihood analyses the
comparison would not be appropriate. For these reasons we define the
predicted correlation length as the distance where the two-point
correlation function is unity.  However, in all cases we also checked
that the values of the slope $\gamma$ obtained from the power-law fit
are well inside the observed $1\sigma$ errorbar. This ensures that the
effect of the different definition of $r_0$ is small.  Finally, we
note that equation \ref{eq:chi2} assumes that the different data are
uncorrelated. This is expected to be a fair assumption, especially for
the X-ray data, which come from different catalogues. In the case of
the optical datasets, some larger correlation is probably present
between the APM samples, due to the presence of common objects in the
two catalogues. However, we will show that we obtain almost identical
conclusions about the confidence levels for the cosmological
parameters considering the APM sample B only or the whole set of
optical catalogues. This shows that the assumption of independence
which underlies our likelihood analysis is not essential for the
results.

Finally, the best-fit cosmological parameters are obtained by
maximizing $\cal L$, i.e. by minimizing $\chi^2$.  The 95.4 and 99.73
per cent confidence levels for the parameters are computed by finding
the region corresponding to an increase $\Delta_{\chi^2}$ with respect
to the minimum value of $\chi^2$: the exact value of $\Delta_{\chi^2}$
depends on the given number of degrees of freedom $\nu_f$.

A similar analysis has been very recently performed by Robinson
(2000), who used a simplified version of the above model and
considered only APM data (all six subsamples). Wherever the comparison
is possible, the results are qualitatively similar, but some
quantitative differences are present. These differences can be
ascribed to a different definition of the effective bias and, above
all, to the absence of any account of the past-light cone effects in
his analysis.

%--------------------------------------------------------
\begin{figure*}
\centering  
\psfig{figure=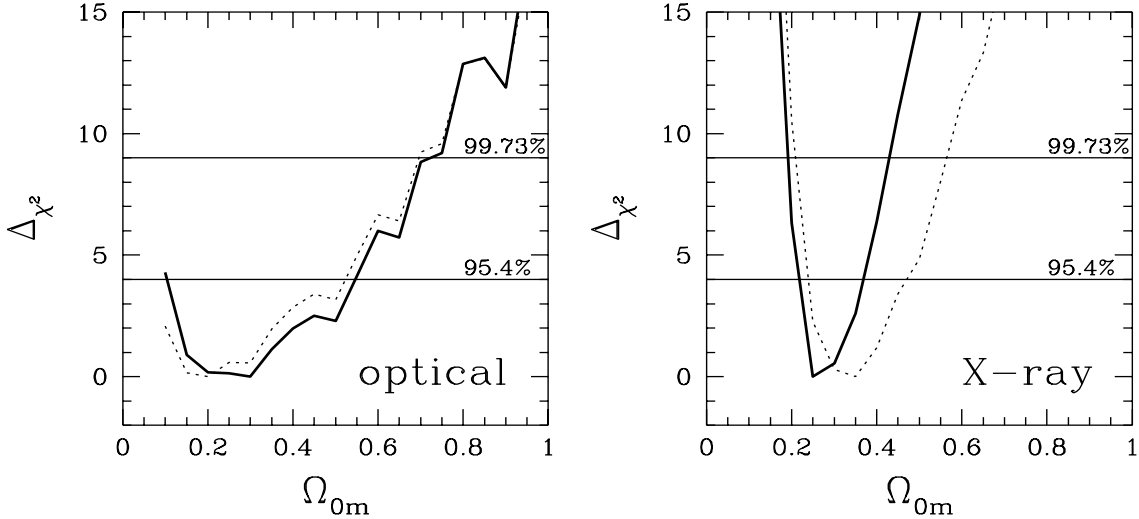,width=16.cm,height=8.cm,angle=0}
\caption{The variation of $\Delta_{\chi^2}$ around the best-fitting 
value of the matter density parameter $\Omega_{\rm 0m}$ for flat CDM
models with shape parameter $\Gamma=0.2$ and normalization reproducing
the cluster abundance.  The left panel refers to the results obtained
using the optical dataset, while the right one refers to the dataset
in the X-ray band. The heavy solid lines correspond to the results
obtained using the complete cluster datasets, the light dotted ones
refer to results obtained using only the largest catalogue (APM sample
B and REFLEX, for optical and X-ray bands, respectively).  Horizontal
lines corresponding to the 95.4 and 99.73 per cent confidence levels
are also shown.}
\label{fi:cont0}
\end{figure*}
%--------------------------------------------------------
%

We start our analysis by considering as a free parameter $\Omega_{\rm
0m}$ only. We fix $\Gamma=0.2$, which is in the range suggested by
different works (see e.g. Peacock \& Dodds 1996), and $\sigma_8$ to
reproduce the cluster abundance. For the normalization we adopt the
fitting formula by Viana \& Liddle (1999), who revised the Henry \&
Arnaud (1991) dataset of cluster X-ray temperatures and included a
treatment of measurement errors. Their formula reads
\be
\sigma_8=0.56 \Omega_{\rm 0m}^{-C}\ , 
\label{eq:sig8}
\ee 
where $C=0.34$ and $C=0.47$ for open and flat models, respectively.
The claimed accuracy of this expression is better than 3 per cent for
$\Omega_{\rm 0m}$ between 0.1 and 1.  The results of our maximum
likelihood analysis obtained by using the complete set of catalogues
are shown for flat models by heavy solid lines in Figure
\ref{fi:cont0}.  The corresponding results for open models (not shown
here) are very similar.  The left panel refer to the analysis of the
optical catalogues.  In this case the minimum of $\chi^2$ is in the
range $\Omega_{\rm 0m}=0.2-0.3$ and we find $\Omega_{\rm 0m}\mincir
0.5$ and $\Omega_{\rm 0m}\mincir 0.7$ at 95.4 and 99.73 per cent
levels, respectively.  The results for the X-ray catalogues are shown
in the right panel.  Also in this case the most likely value for
$\Omega_{\rm 0m}$ is 0.2-0.3 but the allowed regions are much smaller,
i.e. the X-ray data on the correlation length give tighter
constraints: $0.2\mincir \Omega_{\rm 0m}\mincir 0.35$ and $0.2\mincir
\Omega_{\rm 0m}\mincir 0.45$ at 95.4 and 99.73 per cent levels,
respectively.  This result is in qualitative agreement with
our previous analyses, where we found that an Einstein-de Sitter model
can be rejected because it predicts too low a correlation function for
the RASS1 and XBACs catalogues (Moscardini et al. 2000a,b).

In order to understand which subsamples contribute to the constraints,
we repeat the same analysis using in both the optical and X-ray cases
the largest catalogues, i.e. APM sample B and REFLEX. The
corresponding curves are shown in the relative panels as light dotted
lines. Unlike the case of the optical catalogues where very small
differences are found, the constraints obtained using REFLEX only are
less tight: $0.2\mincir \Omega_{\rm 0m}\mincir 0.5$ and $0.2\mincir
\Omega_{\rm 0m}\mincir 0.6$ at 95.4 and 99.73 per cent levels,
respectively.  This shows the importance of including in the analysis
the subsamples at higher X-ray luminosities.

%--------------------------------------------------------
\begin{figure*}
\centering  
\psfig{figure=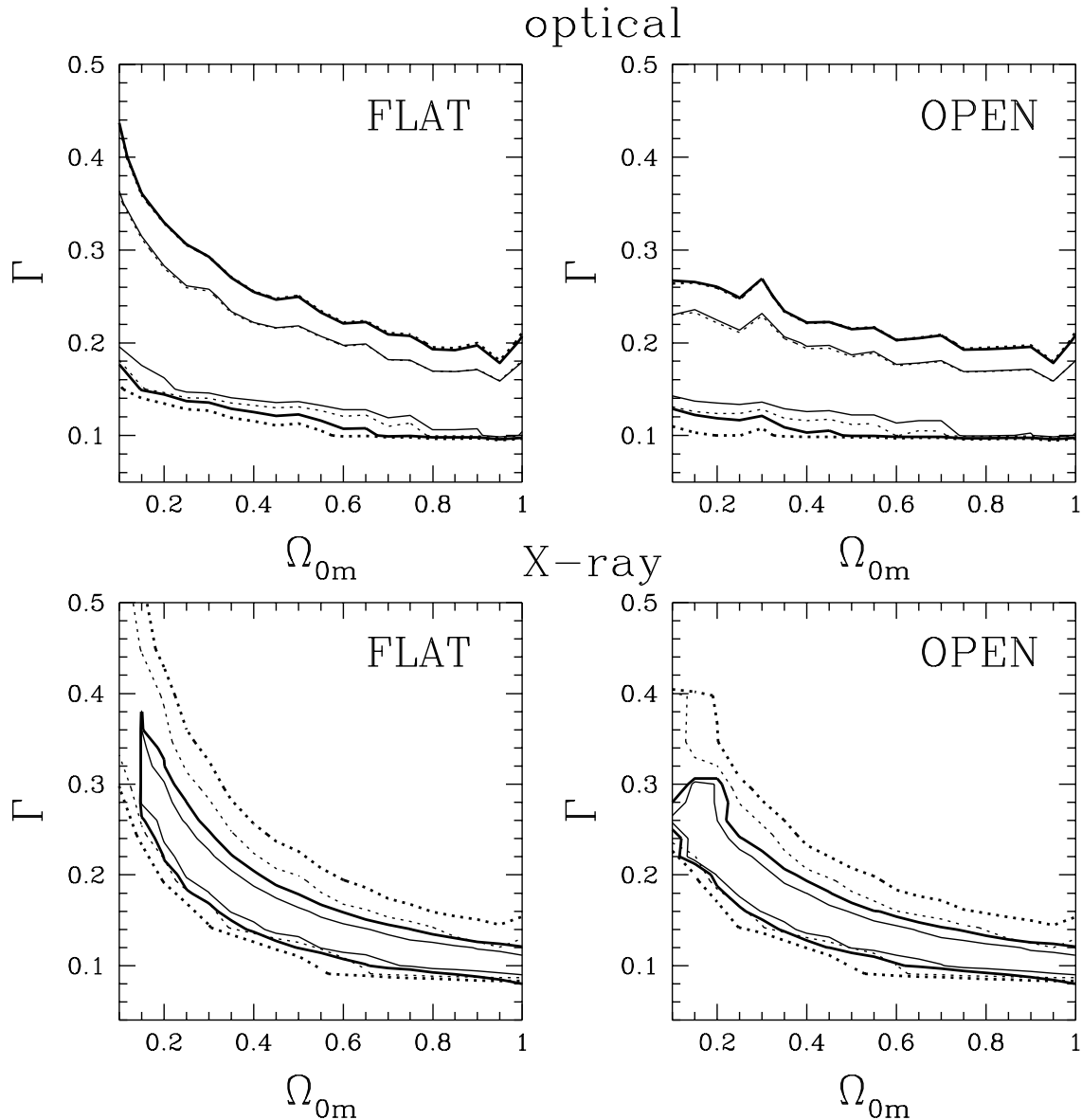,width=16.cm,height=16.cm,angle=0}
\caption{Confidence contours (95.4 and 99.73 per cent confidence levels; 
light and heavy solid lines, respectively) of $\Gamma$ and
$\Omega_{\rm 0m}$ for CDM models with normalization reproducing the
cluster abundance.  The left panels refer to flat cosmological models
with varying cosmological constant $\Omega_{\rm
0\Lambda}=1-\Omega_{\rm 0m}$; the right ones to open models with
vanishing $\Omega_{\rm 0\Lambda}$. The upper row shows the results
obtained using the optical dataset, while the lower one refers to the
dataset in the X-ray band.  The results obtained using only the
largest subsamples (APM sample B for the optical case, REFLEX for the
X-ray band one) are shown by light and heavy dotted lines (95.4 and
99.73 per cent confidence levels, respectively).}
\label{fi:cont1}
\end{figure*}
%--------------------------------------------------------
%

In Figure \ref{fi:cont1} we show the results of our maximum likelihood
analysis when we only fix the model normalisation to reproduce the
cluster abundance, leaving two free parameters: $\Omega_{\rm 0m}$ and
$\Gamma$.

The upper panels refer to the analysis of the optical catalogues.
Confidence levels are shown separately for flat (left panels) and open
(right panels) models. Here $\nu_f=2$, i.e.  $\Delta_{\chi^2}=6.17$
and 11.8, for 95.4 and 99.73 per cent levels, respectively.  The
allowed regions, at least for values of the density parameter
$\Omega_{\rm 0m}\magcir 0.4$, appear to strongly depend only on
$\Gamma$. The data suggest $\Gamma\sim 0.15\pm 0.05$ (errorbars are at
$3\sigma$ level), both for open and flat models.  At smaller
$\Omega_{\rm 0m}$, the resulting values of $\Gamma$ are larger,
especially for models with non-zero cosmological constant. For
instance, for $\Omega_{\rm 0m}=0.2$ we find $\Gamma\sim
0.24^{+0.10}_{-0.08}$ for flat models, and $\Gamma\sim
0.20^{+0.06}_{-0.06}$ for open ones.  The centre of the allowed region
gives a relation between $\Gamma$ and $\Omega_{\rm 0m}$, namely
$\Gamma =0.382-0.797\Omega_{\rm 0m}+1.029
\Omega_{\rm 0m}^2-0.478 \Omega_{\rm 0m}^3$ for the flat models and
$\Gamma =0.215-0.079\Omega_{\rm 0m}-0.064 \Omega_{\rm 0m}^2+0.069
\Omega_{\rm 0m}^3$ for the open models.

The lower panels present the results obtained for the X-ray
catalogues.  Even if the allowed regions are qualitatively compatible
with those displayed by the optical analysis, we notice some
differences. First, the 2- and $3\sigma$ regions are narrower,
i.e. the X-ray data on the correlation length give tighter constraints
on the cosmological parameters. Once again, the $\Omega_{\rm 0m}$
dependence is small for high-density models: we find $\Gamma$ in the
range 0.1-0.15 ($3\sigma$ confidence levels) for $\Omega_{\rm
0m}\magcir 0.5$, with no dependence on the presence of a cosmological
constant. Again, models with low matter density favour larger values
of $\Gamma$, between 0.2 and 0.3. The relations describing the centre
of the allowed region are $\Gamma =0.487-1.342\Omega_{\rm 0m}+1.687
\Omega_{\rm 0m}^2-0.73 \Omega_{\rm 0m}^3$ for the flat models and
$\Gamma =0.394-0.933\Omega_{\rm 0m}+1.084 \Omega_{\rm 0m}^2-0.44
\Omega_{\rm 0m}^3$ for the open models.

Note that, combining the optical and X-ray data altogether, the
allowed regions in the parameter space (not reported in the Figure)
are very similar to those obtained in the analysis of the X-ray
catalogues only, showing once again that these data have larger
discriminating power.

Also in this case we checked the robustness of the results when only
the largest catalogues are used in the analysis. The results,
presented by dotted lines, show that the confidence regions obtained
using the APM sample B are very similar to those obtained using all
three optical catalogues. On the contrary, when we use the REFLEX
catalogue only, the resulting allowed regions, even if compatible with
those obtained using also the catalogues at higher X-ray luminosity,
appear to be larger. For instance, values of $\Gamma$ as large as 0.4
cannot be excluded if the matter density parameter is small
($\Omega_{\rm 0m}\mincir 0.2$).

%
%--------------------------------------------------------
\begin{figure*}
\centering  
\psfig{figure=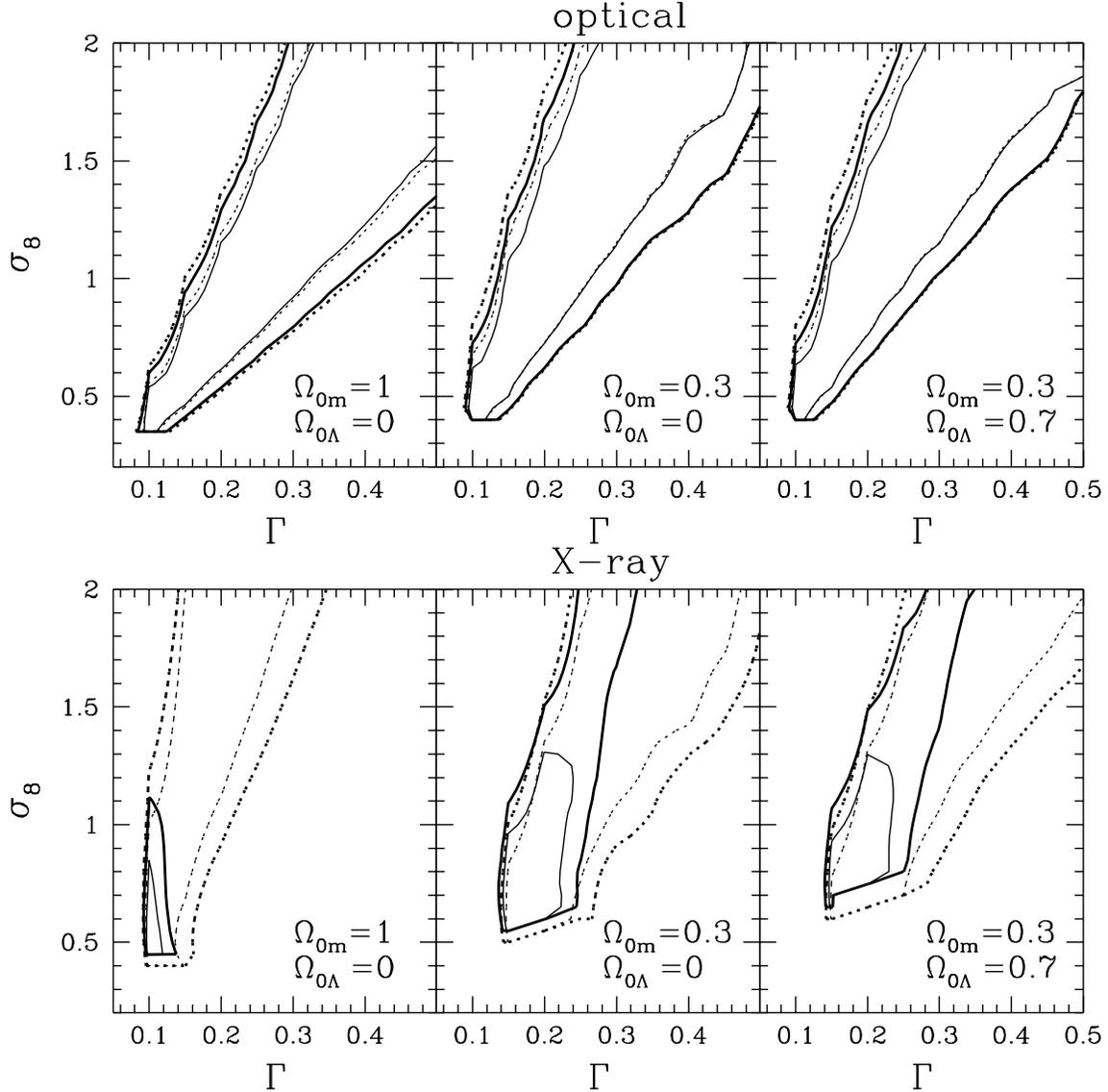,height=16.cm,width=16cm,angle=0}
\caption{Confidence contours (95.4 and 99.73 per cent confidence levels; 
light and heavy solid lines, respectively) of $\Gamma$ and $\sigma_8$
for CDM models with $(\Omega_{\rm 0m},\Omega_{\rm 0\Lambda})=(1,0)$
(left panels), (0.3,0) (central panels) and (0.3,0.7) (right panels).
The upper row shows the results obtained using the optical dataset,
while the lower one refers to the X-ray dataset.  The results obtained
using only the largest subsamples (APM sample B for the optical case,
REFLEX for the X-ray band one) are shown by light and heavy dotted
lines (95.4 and 99.73 per cent confidence levels, respectively).}
\label{fi:cont2}
\end{figure*}
%--------------------------------------------------------

In Figure \ref{fi:cont2} we present the constraints in the
$\Gamma-\sigma_8$ plane, once the values for $\Omega_{\rm 0m}$ and
$\Omega_{\rm 0\Lambda}$ are fixed (i.e. $\nu_f=2$). We consider an
Einstein-de Sitter model, an open model with $\Omega_{\rm 0m}=0.3$ and
a flat model, always with $\Omega_{\rm 0m}=0.3$.  Again the upper
panels refer to the optical data. The confidence limits are quite
similar for the three cosmological models, confirming once again the
weak dependence on the density parameter.  These results can be
directly compared to the analysis made by Robinson (2000; its Figure
3): the agreement is good, even if at a given value of $\sigma_8$
slightly larger values of $\Gamma$ are allowed in our analysis.

The lower panels refer to the X-ray data and show the strongest
constraints coming from these data.  For example, for the Einstein-de
Sitter model only a small region with values of $\Gamma$ quite close
to 0.1 and $0.4\mincir \sigma_8 \mincir 1.1$ is allowed.  The
confidence limits obtained in the case of open and flat models with
$\Omega_{\rm 0m}=0.3$ are similar and are consistent with the analysis
of the optical data, but the allowed region is narrower: the $2\sigma$
region has $\Gamma$ in the range 0.14-0.22 and $\sigma_8$ between 0.6
and 1.3 for the open model, while $0.16\mincir \Gamma \mincir 0.22$
and $0.7\mincir \sigma_8 \mincir 1.3$ for the flat model.
 
Again the combination of optical and X-ray catalogues produces results
almost indistinguishable from those obtained by the X-ray analysis
only. The main difference is the further reduction of the allowed
region in the case of the Einstein-de Sitter model: in this case we
find $\Gamma\sim 0.1$ and $0.45\mincir \sigma_8\mincir0.6$.

When the analysis is limited to the largest catalogues, again no
significant differences are found between the confidence regions
allowed by the APM sample B and the complete set of optical
catalogues. On the contrary, the results for the X-ray catalogues show
that the inclusion in the analysis of the shallower catalogues allows
to reduce the confidence regions, excluding high values of $\Gamma$
and $\sigma_8$.

In conclusions, the previous results show that the constraints coming
from the X-ray datasets are in general tighter than those obtained
from the optical data. This is also evident from the analysis limited
to the largest corresponding catalogues (APM sample B and REFLEX). At
first glance this is unexpected. In fact the errorbars of the
correlation length $r_0$ for the APM sample B are smaller than the
REFLEX ones (see Table \ref{t:data}) and consequently their weights in
the $\chi^2$ estimates are larger. However, the galaxy clusters
belonging to the X-ray datasets typically have a mass larger than the
optical ones. In fact the density of the present optical clusters
requires a minimum mass smaller than that obtained from the limiting
fluxes of the X-ray catalogues here considered. Since the bias factor
entering in the correlation estimates is strongly dependent on the
mass (see, for example, the discussion in Moscardini et al. 2000b),
this difference in mass produces a large spread of the $r_0$ values
predicted for different cosmological models, increasing the
constraining ability of the results. For a similar reason, the
inclusion in our analysis of other X-ray datasets besides REFLEX, even
if with a smaller number of objects and with larger errorbars, helps
in reducing the allowed regions in the parameter space. In fact these
catalogues have a larger limiting flux than REFLEX and sample the
cluster population at higher mass.

\section{Predictions for future surveys}

\subsection{The catalogues}

In this section we will use our model to predict the correlation
length expected in possible future surveys both in the optical and
X-ray bands. Thanks to their depth, in these surveys it would be
possible to obtain information on the high-redshift behaviour of the
cluster two-point function.  For this reason we present our
predictions dividing the data in two different redshift bins ($z\le
0.3$ and $z>0.3$), to allow a discussion of the redshift evolution of
cluster clustering.

In the optical band, a large improvement of our knowledge on the
properties of the large-scale structure as traced by galaxy clusters
will be obtained when the Sloan Digital Sky Survey (SDSS; York et
al. 2000) will be completed.  This survey, which covers an area three
times larger than the APM, will contain redshifts for approximately
one million galaxies. The expected number of galaxy clusters having at
least 100 redshift measurements is approximately 1000. Moreover, the
availability of $5\times 10^7$ galaxies in the photometric data
(complete to a magnitude limit of r'=22) will allow the application of
automated cluster-finding algorithms (e.g. the matched-filter
approach), to extend the cluster catalogue to higher redshifts. In its
final form the SDSS is expected to be nearly as deep as two existing
catalogues, the Palomar Distant Cluster Survey (PDCS; Postman et
al. 1996; Holden et al. 1999) and the ESO Imaging Survey (EIS; Olsen
et al. 1999; Scodeggio et al. 1999), which however are much smaller,
covering only 5.1 and 14.4 square degrees, respectively.  These two
catalogues have been built to find good distant candidates for
successive more detailed observations and cannot be considered
complete and well-defined samples suitable for statistical studies.
Nevertheless, considering the extended versions of these catalogues,
the number density of candidates and the corresponding estimated
redshift distributions are compatible. For these reasons we decided to
use their characteristics to define the properties of a possible
future cluster survey in the optical band. In particular, we use the
number density measured for the EIS catalogue, which consists of 304
objects in the redshift range $0.2\mincir z \mincir 1.3$, with a
median redshift of $z\sim 0.5$. The redshifts have been estimated by
applying the matched-filter algorithm (Postman et al. 1996) and have
an intrinsic uncertainty of at least $\Delta z=0.1$. A large effort
began to validate these cluster candidates. Preliminary results (da
Costa et al. 2000; Ramella et al. 2000) show that more than 65 per
cent of the studied candidates have strong evidence of being real
physical associations. Moreover, first direct spectroscopic
determinations of redshifts are in reasonable agreement with those
derived from the matched-filter algorithm, with a possible systematic
difference of $\Delta z\sim 0.1$. We will compute the predictions of
the clustering properties of this catalogue by fixing the minimum mass
needed to reproduce the EIS cluster density for $z\le 0.3$ and
$z>0.3$.  This choice automatically takes into account the possibility
that false candidates are included in the EIS catalogue by reducing
the required minimum mass. As a consequence, the estimated correlation
function will refer to the objects targeted as candidates and not to
the real clusters which will be actually validated using spectroscopic
techniques.  The underlying assumption is that the failure of the
technique used to select the candidates comes from the inclusion in
the catalogue of high-luminosity systems with a mass just below the
minimum mass of the true EIS clusters.

In the X-ray band, the existing cluster clustering studies were
expected to be largely overcome by the data that the ABRIXAS satellite
(Tr\"umper, Hasinger \& Staubert 1998) was expected to collect,
starting from mid 1999. Unluckily, problems with energy supply caused
the untimely loss of the satellite at the end of April 1999. In the
plans, the ABRIXAS catalogue would have covered an area of 8.27
steradians up to a limiting flux of $S_{\rm lim}=5 \times 10^{-13}$
erg cm$^{-2}$ s$^{-1}$ in the 0.5--2 keV band.  These characteristics
have been used by Moscardini et al. (2000b) to make predictions of the
cluster two-point correlation function.

More recently, it has been proposed to use the very high sensitivity
and good point-spread function of the XMM/Newton satellite,
successfully launched in December 1999, to build a very deep
large-scale structure survey.  The idea is to cover a region of 64
square degrees at high galactic latitude using $24\times 24$ 10ks
XMM/EPIC pointings separated by 20 arcmin offsets.  The expected
limiting flux will be approximately $S_{\rm lim}=5 \times 10^{-15}$
erg cm$^{-2}$ s$^{-1}$ in the 0.5--2 keV band, which is 500 times more
sensitive than the REFLEX one (see Pierre 2000 for more details).  We
predict the clustering properties of this sample using a constant sky
coverage (in absence of real estimates) and assuming the previous
limiting flux.  Hereafter this sample will be called XMM/LSS.

\subsection{Results}
%--------------------------------------------------------
\begin{figure*}
\centering  
\psfig{figure=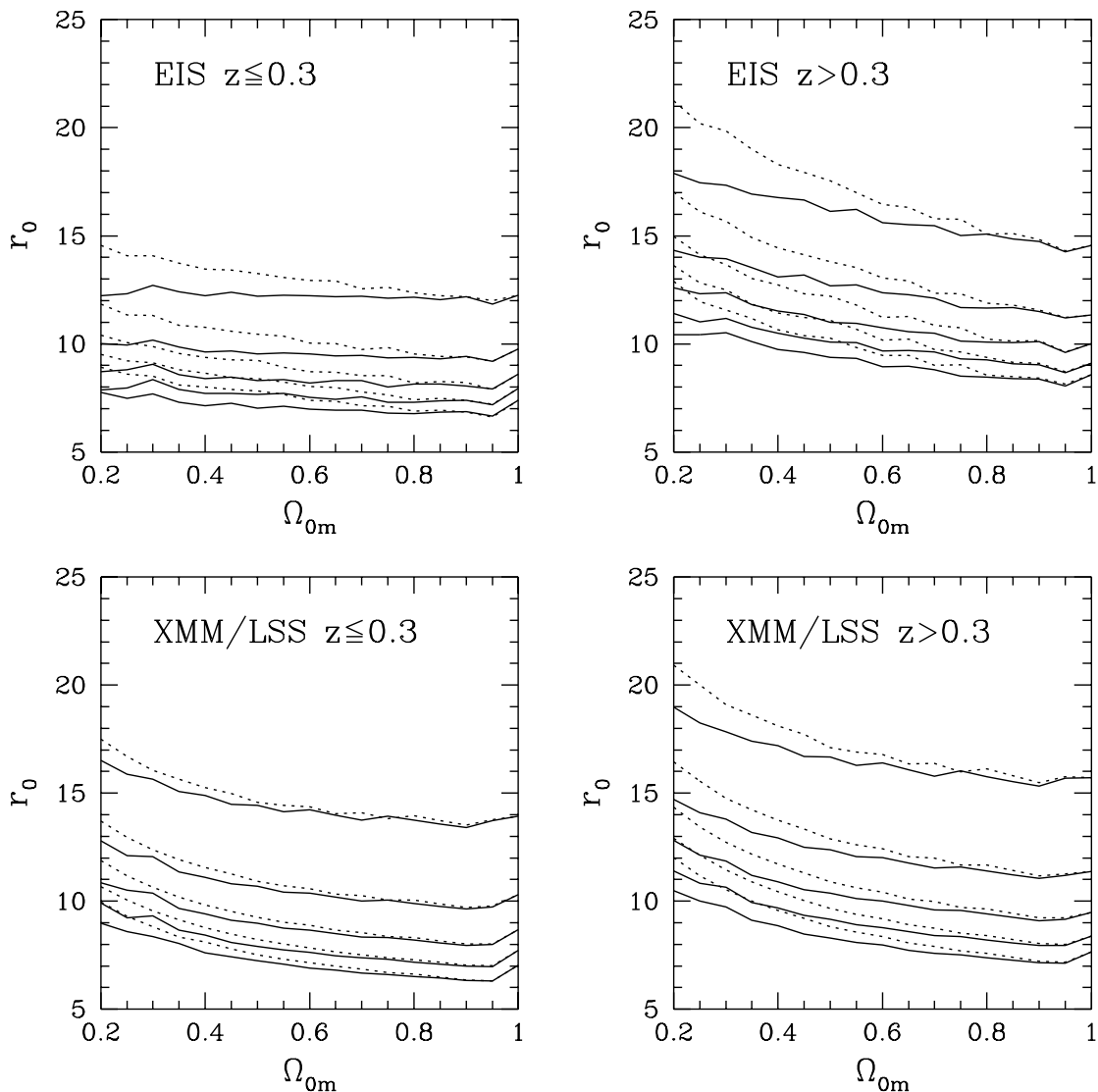,width=16.cm,height=16.cm,angle=0}
\caption{The predicted value of the correlation length (in $h^{-1}$ Mpc)
for different future surveys in the optical band (EIS, left column)
and in the X-ray band (XMM/LSS, left column) as a function of the
present matter density content $\Omega_{\rm 0m}$.  Results are
presented for CDM models with normalization fitted to reproduce the
cluster abundance. The different lines refer to different values of
the $\Gamma$ parameter: 0.1, 0.2, 0.3, 0.4 and 0.5, from top to
bottom.  Dotted lines: flat cosmological models (i.e. with non-zero
cosmological constant); solid ones: open models with vanishing
$\Omega_{\rm 0\Lambda}$. The upper panels show the results obtained
using clusters with $z\le 0.3$, while the lower ones are for $z>0.3$
clusters.}
\label{fi:r0clus}
\end{figure*}
%--------------------------------------------------------
%

In Figure \ref{fi:r0clus} we show the predicted value of the
correlation length for future surveys in the optical band (EIS) and in
the X-ray band (XMM/LSS) as a function of the present matter density
content $\Omega_{\rm 0m}$.  Here we consider only CDM models with
$\sigma_8$ fixed to reproduce the cluster abundances (see equation
\ref{eq:sig8}) and we allow the $\Gamma$ parameter to assume values
from 0.1 to 0.5.  The results are shown separately for clusters having
$z\le 0.3$ (upper panels) and $z>0.3$ (lower panels).  As a general
result, we find that the presence of a cosmological constant increases
the correlation length by a factor always smaller than 15 per
cent. Once again we find that the dependence on the shape parameter
$\Gamma$ is strong: the higher $\Gamma$, the lower the predicted
correlation length. On the contrary varying the matter density
parameter, once $\Gamma$ is fixed, changes $r_0$ only by a factor of
at most 20 per cent. More precisely, we find a slight decrease of the
correlation length with increasing $\Omega_{\rm 0m}$.  All the
catalogues, both in optical and in X-ray bands, display a positive
redshift evolution of the clustering, i.e. the estimates of $r_0$ are
larger for clusters at high redshifts. This result, which confirms a
previous analysis by Moscardini et al. (2000b), is due to the large
increase of the bias factor with redshift; this increase
overcompensates the corresponding decrease of the dark matter
correlation function. In fact, considering high redshifts, galaxy
clusters become rarer objects, connected to higher density
fluctuations.

The relatively low values of the predicted correlation length for the
EIS catalogue show that its cluster number density corresponds to
objects with a small mass, with some possible contamination coming
from false candidates, which would reduce the amplitude of the
expected clustering.

Finally, we notice that the XMM/LSS sample has smaller $r_0$ than the
ABRIXAS survey considered in our previous analysis (Moscardini et
al. 2000b). Moreover the increase of clustering with redshift is less
evident for the XMM/LSS sample than for the ABRIXAS catalogue. This is
what we expect: when the limiting flux is decreased, a large number of
small clusters enter in the catalogue, resulting in a smaller
correlation function (see also Moscardini et al. 2000b).

%
%--------------------------------------------------------
\begin{figure*}
\centering  
\psfig{figure=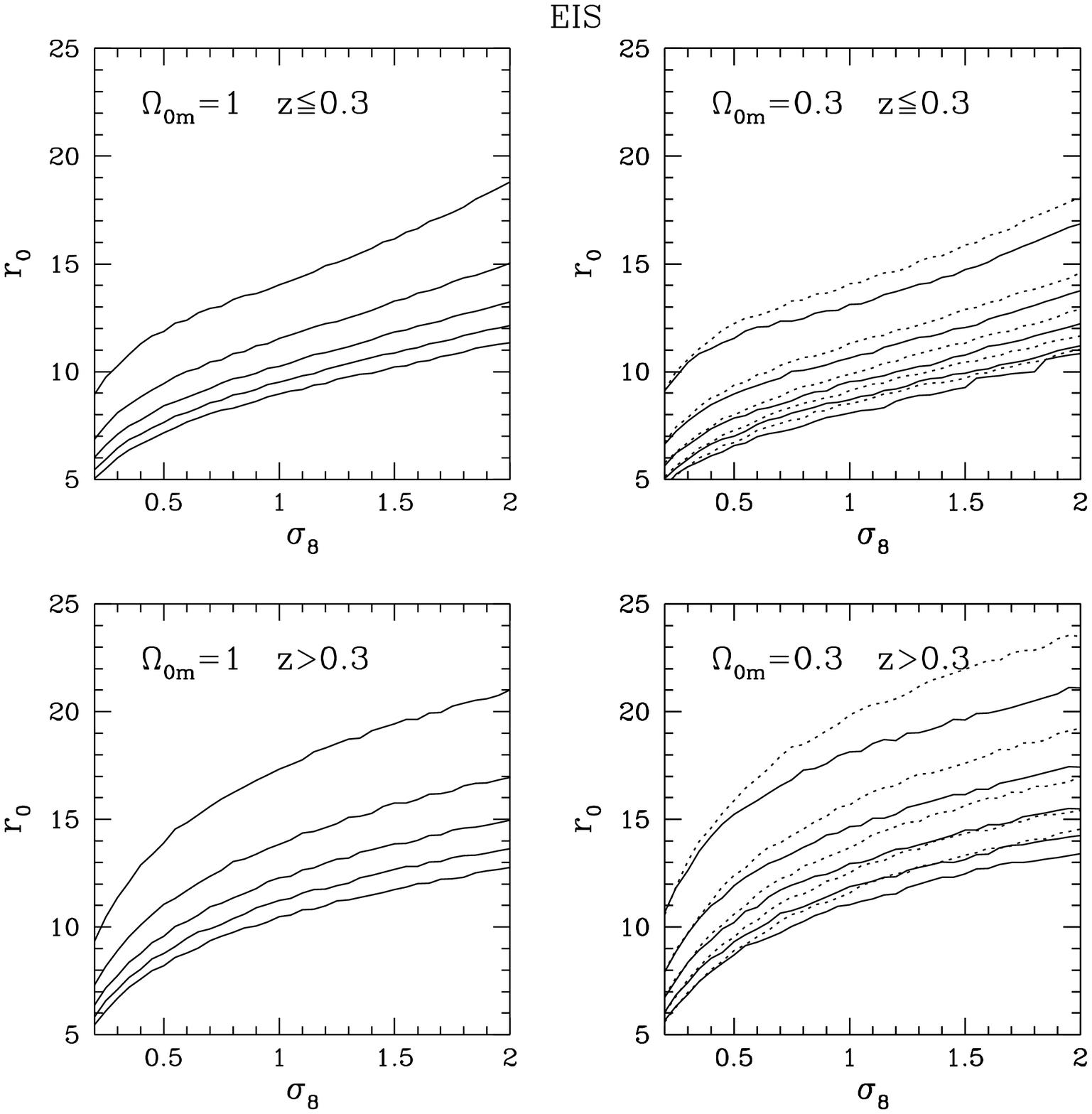,height=16.cm,width=16cm,angle=0}
\caption{The predicted values of the correlation length (in $h^{-1}$ Mpc)
for the EIS survey as a function of the value of the spectrum
normalisation $\sigma_8$.  Results are presented for CDM models with
$(\Omega_{\rm 0m},\Omega_{\rm 0\Lambda})=(1,0)$ (left panels), (0.3,0)
(solid lines in the right panels) and (0.3,0.7) (dotted lines in the
right panels).  The different lines refer to different values of
$\Gamma$: 0.1, 0.2, 0.3, 0.4 and 0.5, from top to down.  The upper
panels shows the results obtained using clusters with $z\le 0.3$,
while the lower one are for $z>0.3$ clusters.}
\label{fi:r0eis}
\end{figure*}
%--------------------------------------------------------
%

Figure \ref{fi:r0eis} shows the predicted correlation length for the
EIS catalogue when the geometry of the universe is fixed and the
values of $\Gamma$ and $\sigma_8$ are varied. We consider here the
Einstein-de Sitter model and the open and flat models with
$\Omega_{\rm 0m}=0.3$.  The small differences in the corresponding
results confirm the slight dependence on the cosmology chosen.  We
find that, at a given $\sigma_8$, $r_0$ is a decreasing function of
$\Gamma$, while, fixing $\Gamma$, $r_0$ is always an increasing
function of $\sigma_8$.

%--------------------------------------------------------
\begin{figure*}
\centering  
\psfig{figure=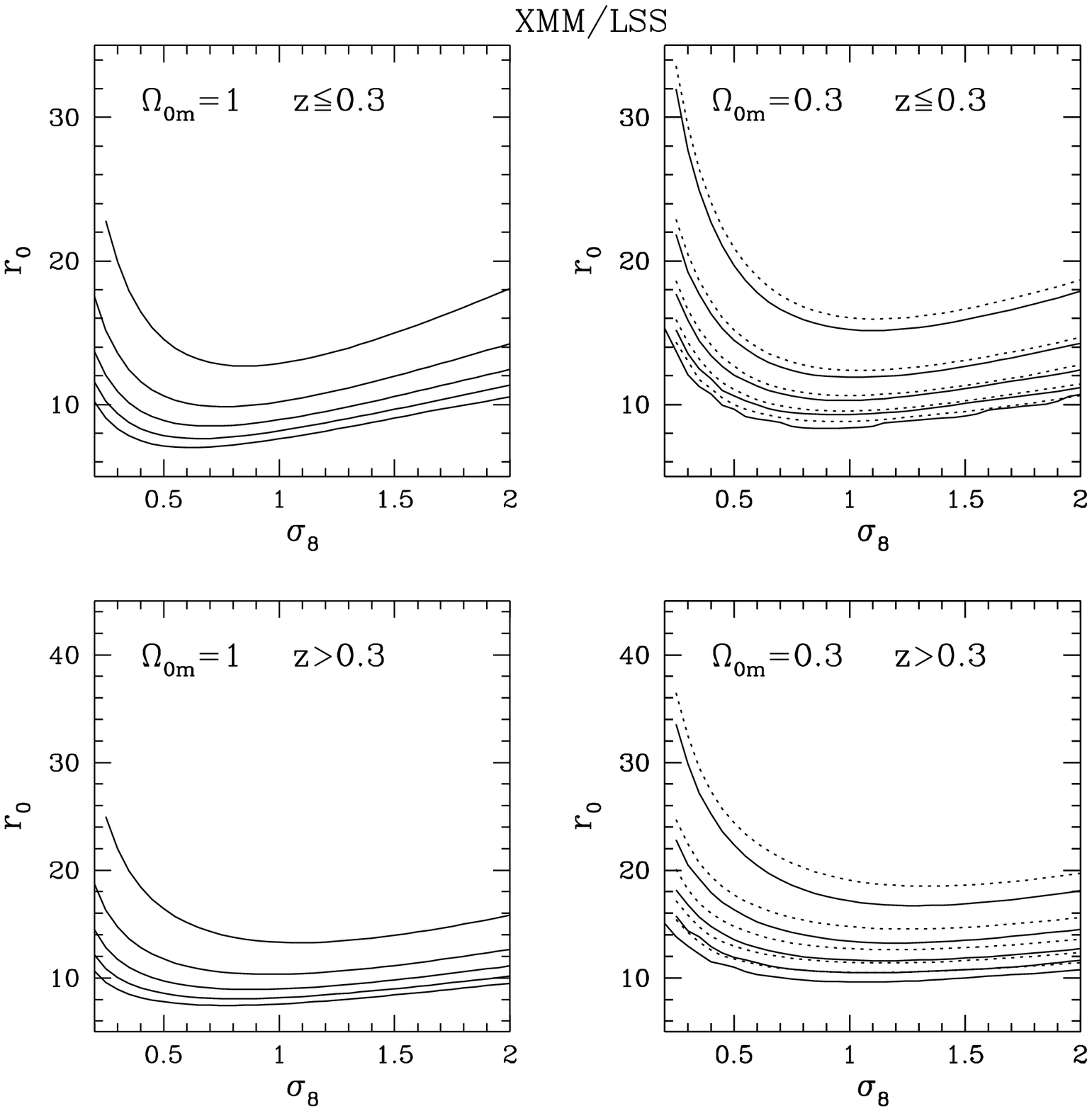,width=16.cm,height=16.cm,angle=0}
\caption{As  Figure~\ref{fi:r0eis} but for the XMM/LSS survey.}
\label{fi:r0xmm}
\end{figure*}
%--------------------------------------------------------
%

A similar analysis has been performed using the characteristics of the
XMM/LSS survey. Comparing the results, shown in Figure
\ref{fi:r0xmm}, with the EIS ones, we can notice that the
$\sigma_8$-dependence of $r_0$ is different: for very small
normalisations, the correlation length is a decreasing function of
$\sigma_8$.  This effect is due to the decrease of the effective bias,
which is more rapid than the growth of the dark matter correlation
function.

%
%--------------------------------------------------------
\begin{figure*}
\centering  
\psfig{figure=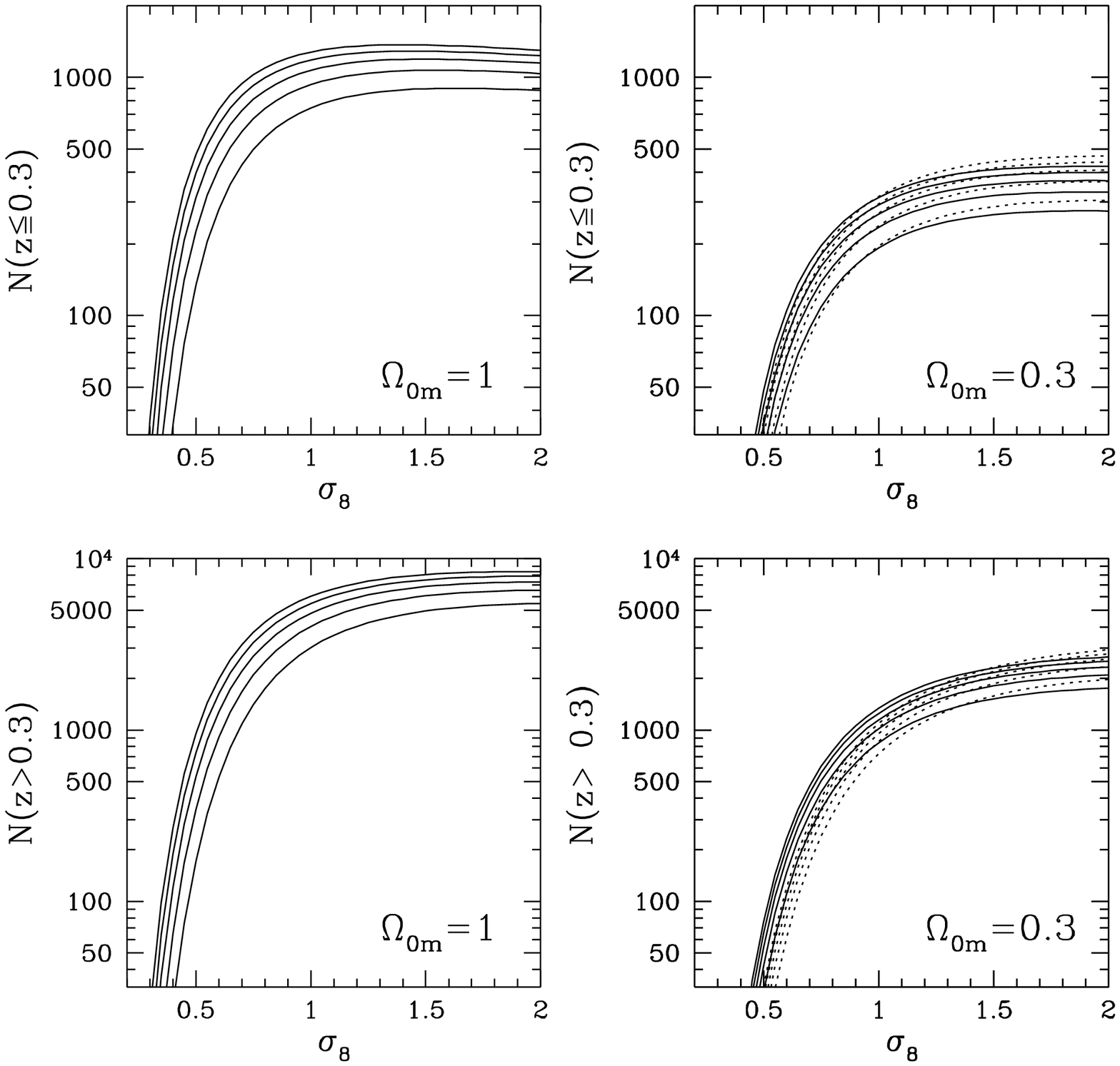,height=16.cm,width=16cm,angle=0}
\caption{The expected number of galaxy clusters
with $z\le 0.3$ (upper panels) and $z>0.3$ (lower panels) in the
XMM/LSS survey as a function of the value of the spectrum
normalisation $\sigma_8$.  Results are presented for CDM models with
$(\Omega_{\rm 0m},\Omega_{\rm 0\Lambda})=(1,0)$ (left panels), (0.3,0)
(solid lines in the right panels) and (0.3,0.7) (dotted lines in the
right panels).  The different lines refer to different values of the
$\Gamma$ parameter: 0.1, 0.2, 0.3, 0.4 and 0.5, from bottom to top.}
\label{fi:nzxmm}
\end{figure*}
%--------------------------------------------------------
%
It is interesting to note that, given a cosmological model, the
predicted values of the correlation length for EIS and XMM/LSS are
quite different, but the spread of $r_0$ obtained in different
cosmological models is expected to be of the same order of magnitude
for the two cases. Unfortunately this does not allow to understand
whether future X-ray or optical surveys will better constrain
cosmological parameters.
 
Another important issue to be discussed is the constraining ability of
the clustering analysis with respect to the standard analysis based on
cluster counts.

In Figure \ref{fi:nzxmm} we show the expected number of galaxy
clusters in the XMM/LSS survey as a function of the spectrum
normalisation $\sigma_8$.  Results refer to the same cosmological
models presented in the previous figures.  We notice that these
numbers are obtained directly from the model described above and do
not pretend to reproduce the observed $\log N-\log S$ relation for
clusters.  To this aim it would be necessary to introduce an ad-hoc
redshift evolution of the temperature-luminosity relation with one
more parameter (see Section 3.2 and the discussion in Moscardini et
al. 2000b).  As already known, the results show a strong dependence on
$\sigma_8$ and a relatively weak dependence on $\Gamma$. For instance,
using the values of the normalisation suggested by equation
\ref{eq:sig8}, the change in the predicted number of clusters for
$0.1\le \Gamma\le 0.3$ (which includes the values suggested by a set
of other observational data, see e.g. Peacock \& Dodds 1996) is at
most 50 per cent.  If we analyse Figure \ref{fi:r0xmm} we find that
with the same assumptions ($\sigma_8$ from cluster abundances and
$0.1\le \Gamma\le 0.3$) the correlation length has a 100 per cent
variation.  This result allows us to conclude that clustering studies
are a good complementary tool to determine the cosmological
parameters.  In particular, once the normalisation $\sigma_8$ is
constrained by other observational data (i.e. cluster counts, cluster
luminosity function or cosmic microwave background), it is quite useful
in fixing the shape parameter $\Gamma$.

\section{Conclusions}

In this paper we discussed the model constraints that it is possible
to infer from the analysis of the observational data on the clustering
of galaxy clusters.  A set of 3 optical (coming from the APM and EDCC
catalogues) and 4 X-ray catalogues (RASS1 Bright Sample, BCS, XBACs,
REFLEX) has been considered. The theoretical predictions for the
different cosmological models have been obtained by using a model
which accounts for the clustering of observable objects in our past
light-cone and for the redshift evolution of both the underlying dark
matter covariance function and the cluster bias factor.  A linear
treatment of redshift-space distortions has been also included. In the
case of X-ray selected clusters we followed the approach of Moscardini
et al. (2000b), which makes use of theoretical and empirical relations
between mass, temperature and X-ray luminosity to convert the limiting
flux of catalogues into a corresponding minimum mass for the dark
matter haloes hosting the clusters.  In the optical band, we used the
mean distance (i.e the number density) of the observed clusters to fix
the minimum mass required by the model.

Our theoretical predictions have been compared with the observed
correlation lengths by means of a maximum-likelihood analysis. We
considered cosmological models belonging to the cold dark matter
class, defined by four parameters: the closure density in dark matter
and vacuum energy ($\Omega_{\rm 0m}$ and $\Omega_{\rm 0\Lambda}$,
respectively); the power-spectrum shape parameter $\Gamma$ and
normalisation $\sigma_8$.

Our main results can be summarized as follows: 

\begin{itemize} 
\item
The constraints coming from X-ray and optical clusters are consistent
but the former appear to be tighter.
\item 
If we fix the power-spectrum shape parameter $\Gamma=0.2$, in the
range suggested by different observational datasets, and the
power-spectrum normalisation to reproduce the cluster abundance, we
obtain strong constraints on the value of the matter density
parameter, independently of the presence of a cosmological constant:
$\Omega_{\rm 0m}\le 0.5$ and $0.2\le \Omega_{\rm 0m}\le 0.35$ at the
$2\sigma$ level, for the optical and X-ray data, respectively.
\item
If we allow the shape parameter to vary, we find that the clustering
properties of clusters are only weakly dependent on $\Omega_{\rm 0m}$
and $\Omega_{\rm 0\Lambda}$.  On the contrary, the results appear to
be much more strongly sensitive to $\Gamma$. In fact, smaller shape
parameters correspond to higher correlation lengths.
\item
Considering models with the normalisation coming from the cluster
abundance, we find that the centre of the region allowed by the
maximum-likelihood analysis of the optical data is described by the
relation $\Gamma =0.382-0.797\Omega_{\rm 0m}+1.029
\Omega_{\rm 0m}^2-0.478 \Omega_{\rm 0m}^3$ and
$\Gamma =0.215-0.079\Omega_{\rm 0m}-0.064 \Omega_{\rm 0m}^2+0.069
\Omega_{\rm 0m}^3$ for flat and open models, respectively.
Considering the catalogues in the X-ray band, we find $\Gamma
=0.487-1.342\Omega_{\rm 0m}+1.687 \Omega_{\rm 0m}^2-0.73 \Omega_{\rm
0m}^3$ for the flat models and $\Gamma =0.394-0.933\Omega_{\rm
0m}+1.084 \Omega_{\rm 0m}^2-0.44 \Omega_{\rm 0m}^3$ for the open
models.
\item 
Using X-ray selected data only, we find that for the Einstein-de
Sitter model the value of $\Gamma$ has to be quite close to 0.1 with
$0.4\mincir \sigma_8 \mincir 1.1$; for open and flat models with
$\Omega_{\rm 0m}=0.3$ the $2\sigma$ region has $0.14 \mincir \Gamma
\mincir 0.22$ and $0.6 \mincir \sigma_8 \mincir 1.3$.
\end{itemize}

We also used our model to make predictions on the clustering
properties of galaxy clusters expected in future surveys. We
considered an optical catalogue with characteristics similar to the
EIS project and a deep X-ray catalogue with the characteristics of the
XMM/LSS survey. From this analysis we can conclude that:

\begin{itemize} 
\item
Clusters at high redshifts are expected to have a larger correlation
function than at low redshifts.
\item
Again, predictions are almost insensitive to the presence of a
cosmological constant while they are strongly dependent on the shape
parameter.
\item
The predicted clustering for the EIS catalogue is relatively small for
all cosmological models suggesting that its cluster number density
corresponds to objects with small mass, including some possible false
candidates.
\item 
The correlation length for X-ray selected clusters is confirmed to
depend on the limiting flux of the survey: the deeper the catalogue,
the smaller $r_0$. Moreover, the redshift evolution of clustering is
less evident in deeper catalogues.
\end{itemize}

In conclusion, our results show that the existing data on the
clustering properties of clusters can be successfully used to put
constraints on the cosmological parameters.  The future availability
of deeper surveys can increase the power of this approach, which can
be considered complementary to the traditional study of cluster
abundances.

\section*{Acknowledgments.} 

This work has been partially supported by Italian MURST, CNR and ASI,
and by the TMR european network ``The Formation and Evolution of
Galaxies" under contract n. ERBFMRX-CT96-086. LM thank the
Max-Planck-Institut f\"ur Astrophysik for its hospitality during the
visit when this work was completed.  We want to warmly thank Sabrina
De Grandi for stimulating comments and helpful suggestions. We are
grateful to C. Collins for having provided the clustering results of
the REFLEX catalogue before publication and to Herv\'e Aussel, Marco
Scodeggio and Bepi Tormen for clarifying discussions.  We thank the
anonymous referee for useful suggestions which helped to improve the
presentation of our results.

\end{document}